\documentclass[12pt]{article}
\usepackage{amssymb,graphicx}

\def\np{\vspace{12pt} \noindent}

\def\arcmin{\hbox{$^\prime$}}
\def\arcsec{\hbox{$^{\prime\prime}$}}

\begin{document}

\begin{center}
{\LARGE \bf {A large population of 'Lyman-break' galaxies in a protocluster at redshift $z\approx4.1$}}
\end{center}
\begin{center}
{\large \bf {George K. Miley$^1$, Roderik A. Overzier$^1$, Zlatan I. Tsvetanov$^2$, Rychard J. Bouwens$^3$, Narciso Ben\'\i tez$^2$, John P. Blakeslee$^2$, Holland C. Ford$^2$, Garth D. Illingworth$^3$, Marc Postman$^4$, Piero Rosati$^5$, Mark Clampin$^4$, George F. Hartig$^4$, Andrew W. Zirm$^1$, Huub J. A. R\"ottgering$^1$, Bram P. Venemans$^1$, David R. Ardila$^2$, Frank Bartko$^6$, Tom J. Broadhurst$^7$, Robert A. Brown$^2$, Chris J. Burrows$^2$, E. S. Cheng$^8$, Nicholas J. G. Cross$^2$, Carlos De Breuck$^5$, Paul D. Feldman$^2$, Marijn Franx$^1$, David A. Golimowski$^2$, Caryl Gronwall$^2$, Leopoldo Infante$^9$, Andr\'e R. Martel$^2$, Felipe Menanteau$^2$, Gerhardt R. Meurer$^2$, Marco Sirianni$^2$, Randy A. Kimble$^8$, John E. Krist$^6$, William B. Sparks$^4$, Hien D. Tran$^2$, Richard L. White$^4$ \& Wei Zheng$^2$}}\\

\medskip
\noindent
\begin{small}
$^1$ Leiden Observatory, University of Leiden, PO Box 9513, Leiden, 2300 RA, The Netherlands \\
$^2$ Department of Physics \& Astronomy, Johns Hopkins University, Baltimore, Maryland 21218, USA \\
$^3$ Lick Observatory, University of California, Santa Cruz, California 95064, USA \\
$^4$ Space Telescope Science Institute, Baltimore, Maryland 21218, USA \\
$^5$ European Southern Observatory, Garching, D-85748, Germany \\
$^6$ Bartko Science \& Technology, Mead, Colorado 80542-0670, USA \\
$^7$ The Racah Institute of Physics, Hebrew University, Jerusalem, 91904, Israel \\
$^8$ NASA-Goddard Space Flight Centre, Greenbelt, Maryland 20771, USA \\
$^9$ Departmento de Astronomia y Astrofisica, Pontificia Universidad Catolica de Chile, Casilla 306, Santiago 22, Chile\\
\end{small}
\end{center}

{\np \bf
The most massive galaxies and the richest clusters are believed to have emerged from regions with the largest enhancements of mass density$^{1-­4}$ relative to the surrounding space. Distant radio galaxies may pinpoint the locations of the ancestors of rich clusters, because they are massive systems associated with overdensities of galaxies that are bright in the Lyman-$\alpha$ line of hydrogen$^{5-­7}$. A powerful technique for detecting high-redshift galaxies is to search for the characteristic `Lyman break' feature in the galaxy colour, at wavelengths just shortwards of Ly$\alpha$, due to absorption of radiation from the galaxy by the intervening galactic medium. Here we report multicolour imaging of the most distant candidate$^{7-­9}$ protocluster, TN J1338­-1942 at a redshift $z\approx4.1$. We find a large number of objects with the characteristic colours of galaxies at that redshift, and we show that this excess is concentrated around the targeted dominant radio galaxy. Our data therefore indicate that TN J1338-­1942 is indeed the most distant cluster progenitor of a rich local cluster, and that galaxy clusters began forming when the Universe was only 10 per cent of its present age.
}

\np
There is increasing evidence that structures of galaxies existed in the early Universe, but the detection of protoclusters at redshifts $z>1$ using conventional optical and X-ray techniques is difficult$^{10-­12}$. Some of us have developed an efficient method for pinpointing distant protoclusters. The technique is based on the hypothesis that the most powerful known high-redshift radio galaxies are frequently associated with massive forming galaxies$^{13­-16}$ in protoclusters$^5$. As a first step towards testing this hypothesis, we recently conducted a large programme with the Very Large Telescope (VLT) of the European Southern Observatory to search for galaxy overdensities associated with protoclusters around luminous high-redshift radio galaxies. Deep narrow- and broad-band imaging was used to locate candidate galaxies having bright Ly$\alpha$ emission, and follow-up spectra have confirmed that most of these candidates have similar redshifts to the high-redshift radio galaxies. All five targets studied with the VLT to sufficient depth have $>20$ spectroscopically confirmed Ly$\alpha$ and/or H$\alpha$ companion galaxies, associated with galaxy overdensities$^{6,7}$. Their formal velocity dispersions are a few hundred km s$^{-1}$, but there was not enough time since the Big Bang for them to have become virialized. The scale sizes of the structures inferred from their spatial boundaries are $\sim3-­5$ Mpc. Assuming that the overdensities are due to a single structure, the masses derived from the observed structure sizes and overdensities are comparable to those of clusters of galaxies in the local Universe$^7$. These observations led us to hypothesize that the overdensities of Ly$\alpha$ galaxies around radio sources are due to the fact that they are in protoclusters.

\np
Galaxies that emit strong Ly$\alpha$ comprise only a small fraction of distant galaxies, and are biased towards non-dusty objects and galaxies that are undergoing the most vigorous star formation. Only about 25\% of $z\approx3$ galaxies have Ly$\alpha$ equivalent widths detectable by our VLT narrow-band imaging searches$^{17,18}$. If the overdensities of Ly$\alpha$ galaxies are located in protoclusters, additional galaxy populations should be present and detectable on the basis of characteristic continuum features in the galaxy spectra. The most important of these features is the sharp `Lyman break' blueward of Ly$\alpha$, caused by the absorption of the galaxy continuum radiation by neutral hydrogen clouds along the line of sight. Searching for Lyman-break galaxies is a powerful technique for finding highredshift galaxies$^{11,19,20}$.

\np
Because of its high spatial resolution, large field of view and excellent sensitivity, the Advanced Camera for Surveys$^{21}$ (ACS) on the Hubble Space Telescope is uniquely suited for studying the morphologies of galaxies in the protoclusters and for finding additional galaxies on the basis of the Lyman-break features in their spectra. We therefore used the ACS to observe the most distant of our VLT protoclusters, TN J1338­-1942 (ref. 7) at $z=4.1$. This is a structure with 21 spectroscopically confirmed Ly$\alpha$ emitters and a rest-frame velocity dispersion of 325 km s$^{-1}$. Images were taken through three `Sloan' filters--g band centred at 4,750 \AA, r band centred near 6,250 \AA\ and i band centred near 7,750 \AA. These filters were chosen so that their wavelength responses bracketed redshifted Ly$\alpha$ at 6,214 \AA\ and were sensitive to the Lyman-break feature blueward of Ly$\alpha$.

\np
A 3.4\arcmin$\times$3.4\arcmin\ field was observed, with the radio galaxy located $\sim1\arcmin$ from the image centre. Besides the radio galaxy, this field covered 12 of the 21 known Ly$\alpha$ emitting galaxies in the candidate protocluster. All 12 objects were detected in both r band and i band, with i-band magnitudes ranging from 25 to 28, compared with $23.3\pm0.03$ for the radio galaxy. As illustrated in Fig. 1, these objects were either absent or substantially attenuated in the g band, and their $g-­r$ colours are generally consistent with predicted values of Lyman breaks$^{22}$. Half of the objects are extended in $i_{775}$, and three of these are resolved into two distinct knots of continuum emission, suggestive of merging.

\np
We next used the Lyman-break technique to search for a population of Lyman-break galaxies in the protocluster that do not emit strong Ly$\alpha$ and would therefore have been undetectable in our VLT observations. Evidence for the existence of such a population was sought by analysing the number and spatial distribution of `g-band dropout' objects--that is, objects whose colours are consistent with Lyman breaks in their spectra at the redshift of the protocluster. To investigate whether there is a statistically significant excess of such g-band dropout objects, we estimated the surface density and cosmic variance of g-band dropouts in a typical ACS field observed with the same filters and to the same depth as TN J1338-­1942. We did this by cloning$^{23}$ $B_{435}$-band dropouts in 15 different pointings from the southern field of the Great Observatories Origins Deep Survey (GOODS)$^{24}$. Results indicate that the number of g-band dropouts in our field is a factor of 2.5 times higher than the average number found in a random GOODS field. Taking account of the typical cosmic variance$^{25}$ in the distribution of $z\approx4$ Lyman-break galaxies, this is a $3\sigma$ excess on the assumption that the distribution function is gaussian. Further evidence that a substantial fraction of these g-dropout objects are Lyman-break galaxies associated with the protocluster is provided by the strong concentration of the g-band dropouts in a cluster-sized region around the radio galaxy. This is illustrated in Fig. 2. More than half of the g-band dropouts are located in a region of $\sim1\arcmin$ in radius (corresponding to a diameter of $\sim1$ Mpc at $z=4.1$). The number of g-band dropouts in this region is a factor of 5 times the average number encountered in similarly sized regions that are randomly drawn from the GOODS survey. This is a $5\sigma$ excess, indicating that the number of g-band dropouts in our field is anomalously high at greater than the 99\% confidence level. The spatial non-uniformity of g-band dropout objects in our field becomes even more pronounced when fainter objects down to a magnitude of $=27$ are included (Fig. 2).

\np
Are there alternative explanations for the observed excess of g-band dropout objects other than a population of Lyman-break galaxies at $z\approx4.1$? An object with a Balmer break at $z\approx0.5$ could also be observed as a g-band dropout object. However, a population of such $z\approx0.5$ objects would also be present in the GOODS comparison sample. Although the existence of an intervening structure of Balmer-break galaxies at $z\approx0.5$ cannot be completely ruled out, its coincidence in location with the $z\approx4.1$ structure of Ly$\alpha$ galaxies and the faintness and small sizes of the observed objects make this possibility highly unlikely. 

\np
The spatial coincidence of the excess in g-band dropout objects with the previously detected overdensity of Ly$\alpha$ emitters around a forming massive galaxy is strong evidence that we are observing a new population of Lyman-break galaxies in a protocluster. This would mean that TN J1338-­1942, at $z\approx4.1$, is indeed the most distant known protocluster, and that distant luminous radio galaxies pinpoint the progenitors of nearby rich clusters. Such protoclusters provide an opportunity to study the development of galaxies and clusters in the early Universe. They provide samples of different galaxy populations at the same distance, whose morphologies and spectral energy distributions could be used to disentangle the evolution and star formation history of different types of galaxies. The topological information that could be derived by mapping the shapes and sizes of such protoclusters over larger areas could answer the question of whether the first protoclusters in the early Universe formed in sheets or filaments.

\np
1. Kaiser, N. On the spatial correlation function of Abell clusters. Astrophys. J. 284, L9-­L12 (1984).\\
2. White, S. D. M. \& Rees, M. J. Core condensation in heavy halos--A two-stage theory for galaxy formation and clustering. Mon. Not. R. Astron. Soc. 183, 341­-358 (1978). \\
3. Baugh, C. M., Cole, S., Frenk, C. S. \& Lacey, C. G. The epoch of galaxy formation. Astrophys. J. 498, 504-­521 (1998). \\
4. Bahcall, N. A. \& Fan, X. The most massive distant clusters: Determining $\Omega$ and $\sigma_8$. Astrophys. J. 504, 1­6 (1998). \\
5. Miley, G. in Extrasolar Planets to Cosmology: The VLT Opening Symposium (eds Bergeron, J. \& Renzini, A.) 32-42 (Springer, Berlin, 2000). \\
6. Pentericci, L. et al. A search for clusters at high redshift. II. A proto cluster around a radio galaxy at $z=2.16$. Astron. Astrophys. 361, L25-­L28 (2000). \\
7. Venemans, B. P. et al. The most distant structure of galaxies known: A protocluster at $z=4.1$. Astrophys. J. 569, L11­-L14 (2002). \\
8. De Breuck, C. et al. VLT spectroscopy of the $z=4.11$ radio galaxy TN J1338-­1942. Astron. Astrophys. 352, L51-­L56 (1999). \\
9. De Breuck, C., van Breugel, W., Rottgering, H. J. A. \& Miley, G. A sample of 669 ultra steep spectrum radio sources to find high redshift radio galaxies. Astron. Astrophys. Suppl. 143, 303-­333 (2000). \\
10. Rosati, P. et al. An X-ray-selected galaxy cluster at $z=1.26$. Astron. J. 118, 76-­85 (1999). \\
11. Steidel, C. C. et al. A large structure of galaxies at redshift $z\sim3$ and its cosmological implications. Astrophys. J. 492, 428­438 (1998). \\
12. Shimasaku, K. et al. Subaru deep survey IV: Discovery of a large-scale structure at redshift $\sim=5$. Astrophys. J. 586, L111-­L114 (2003). \\
13. De Breuck, C. et al. Optical and near-infrared imaging of ultra-steep-spectrum radio sources: The K-z diagram of radio-selected and optically selected galaxies. Astron. J. 123, 637-­677 (2002). \\
14. Dey, A., van Breugel, W., Vacca, W. D. \& Antonucci, R. Triggered star formation in a massive galaxy at $z=3.8$: 4C 41.17. Astrophys. J. 490, 698-­709 (1997). \\
15. Pentericci, L. et al. HST images and properties of the most distant radio galaxies. Astrophys. J. 504, 139-­146 (1999). \\
16. van Ojik, R. {\it Gas in Distant Radio Galaxies: Probing the Early Universe}. Thesis, Leiden Univ. (1995). \\
17. Steidel, C. C. et al. Lya imaging of a proto-cluster region at $\left<z\right>=3.09$. Astrophys. J. 532, 170-­182 (2000). \\
18. Shapley, A. E., Steidel, C. C., Pettini, M. \& Adelberger, K. Rest-frame ultraviolet spectra of $z\sim3$ Lyman break galaxies. Astrophys. J. 588, 65-­89 (2003). \\
19. Steidel, C. C., Giavalisco, M., Pettini, M., Dickinson, M. \& Adelberger, K. Spectroscopic confirmation of a population of normal star-forming galaxies at redshifts $z\>3$. Astrophys. J. 462, L1-­L7 (1999). \\
20. Steidel, C. C., Adelberger, K., Giavalisco, M., Dickinson, M. \& Pettini, M. Lyman-break galaxies at $z>4$ and the evolution of the ultraviolet luminosity density at high redshift. Astrophys. J. 519, 1­-17 (1999). \\
21. Ford, H. C. et al. Advanced camera for the Hubble Space Telescope. Proc. SPIE 3356, 234-­248 (1998). \\
22. Madau, P. Radiative transfer in a clumpy universe: The colors of high-redshift galaxies. Astrophys. J. 441, 18-­27 (1995). \\
23. Bouwens, R. J., Broadhurst, T. \& Illingworth, G. Cloning dropouts: Implications for galaxy evolution at high redshift. Astrophys. J. 593, 640--660 (2003). \\
24. Giavalisco, M. et al. The Great Observatories Origins Deep Survey. Astrophys. J. Lett. (special issue) (in the press). \\
25. Somerville, R. S. et al. Cosmic variance in the Great Observatories Origins Deep Survey. Astrophys. J. Lett. (special issue) (in the press). \\
26. Blakeslee, J. P., Anderson, K. R., Meurer, G. R., Ben\'\i tez, N. \& Magee, D. An automatic image reduction pipeline for the Advanced Camera for Surveys. ASP Conf. Ser. 295, 257--260 (2003). 
27. Bertin, E. \& Arnouts, S. SExtractor: Software for source extraction. Astron. Astrophys. 117, 393-­404 (1996).\\

\np
{\small{Correspondence and requests for materials should be addressed to G.K.M. (miley@strw.leidenuniv.nl).}}

\pagebreak

\np
{\bf Figure 1}: Deep images of Ly$\alpha$-emitting protocluster galaxies. Images show galaxy morphologies observed through three filters: g band (left), r band (middle) and i band (right). Each $2.5\arcsec\times2.5\arcsec$ image has been smoothed by a gaussian function with a fullwidth at half-maximum 
of 1.5 pixels ($0.074\arcsec$). The observations were carried out between 8 and 12 July 2002 with the Wide Field Channel of the ACS$^{21}$. The total observing time of 13 orbits was split over the broad-band filters F475W (g band, four orbits), F625W (r band, four orbits) and F775W (i band, five orbits), thereby bracketing redshifted Ly$\alpha$ at 6,214 \AA. During each orbit, two 1,200-s exposures were made, to facilitate the removal of cosmic rays. The observations were processed through the ACS GTO pipeline$^{26}$ to produce registered, cosmic-ray-rejected images. The limiting $2\sigma$ magnitudes in a 0.2-arcsec$^2$ aperture were 28.71 (F475W), 28.44 (F625W) and 28.26 (F775W). Object detection and photometry were then obtained using SExtractor$^{27}$. {\bf a}, The clumpy radio galaxy TN J1338-­1942 at $z=4.1$. This is the brightest galaxy in the protocluster and inferred to be the dominant cluster galaxy in the process of formation. Because the equivalent width of Ly$\alpha$ is large ($\sim500$ \AA), the r band is dominated by Ly$\alpha$. Arrows indicate the positions$^{8}$ of the radio core (C) and the northern hotspot (H). The Ly$\alpha$ emission is elongated in the direction of the radio emission and the large-scale Ly$\alpha$ halo${^7}$ with a projected linear size of $\sim15$ kpc (assuming $H_0=65$ km s$^{-1}$ Mpc$^{-1}$, $\Omega_M=0.3$, $\Omega_M=0.7$). {\bf b}, Images of five spectroscopically confirmed Ly$\alpha$ emitters in the protocluster. Listed below each galaxy are its spectroscopic redshift$^7$, the magnitude of the observed Lyman break, and the i-band magnitude. Two of the Ly$\alpha$ emitters are clumpy, as expected from young galaxies.

\np
{\bf Figure 2}: The spatial distribution of g-band dropout objects. Superimposed on the combined $3.4\arcmin\times3.4\arcmin$ ACS greyscale image are the locations of g-band dropout objects (blue circles), selected to have colours and magnitudes of $(g-­r)\ge1.5$, $(g-­r)\ge(r-­i)+1.1$, $(r-­i)\le1$ and $i<27$. In addition, objects were required to have a SExtractor$^{27}$ stellarity parameter of less than 0.85 to ensure that the sample was not contaminated by stars. These criteria filter galaxies having Lyman breaks at $z\approx 4$, thereby providing a sample of protocluster Lyman-break galaxy candidates. We detected 30 g-dropout objects in the field around TN J1338-­1942 with $i_{775}<26$, and 56 with $i_{ 775}<27$. The number of g-band dropout objects is anomalously large, and their distribution is concentrated within a circular region of $\sim1\arcmin$ in radius that includes the radio galaxy TN J1338-­1942 (large green circle). Also shown are the positions of the spectroscopically confirmed Ly$\alpha$ emitters (red squares). Because the selection criteria were optimized to detect Lyman-break galaxies, some of the Ly$\alpha$ emitters did not fall into the formal sample of Lyman-break galaxies. The measured excess and its spatial clustering are evidence that a substantial fraction of the g-band dropout objects are Lyman-break galaxies associated with the protocluster. (See Fig. 1 legend for further details about the observations and the subsequent analysis.) Scale bar, 1\arcmin.





\end{document}